\documentclass[preprint,showpacs,floatfix,preprintnumbers,amsmath,amssymb]{revtex4}
\usepackage{graphicx}
\usepackage{dcolumn}
\usepackage{bm}

\begin{document}

\title{Group-V mixing effects in the 
structural and optical properties of $(ZnSi)_{1/2}P_{1/4}As_{3/4}$}

\author{A. A. Leit\~ao $^{1,2}$, R. B. Capaz $^1$}
\affiliation{$^1$ Instituto de F\'\i sica, Universidade Federal do Rio de Janeiro, Caixa Postal 68528, 21941-972, Rio de Janeiro, RJ, Brazil \\
$^2$ Departamento de Qu\'\i mica,
Universidade Federal de Juiz de Fora, CEP 36036-330, Juiz de Fora, MG,
Brazil (leitao@quimica.ufjf.br)
}

\date{\today}

\begin{abstract}
We present {\it ab initio} total energy and band structure calculations based on
Density Funtional Theory (DFT)
within the Local Density Aproximation (LDA) on group-V mixing effects in the
 optoelectronic material    
$(ZnSi)_{1/2}P_{1/4}As_{3/4}$. This compound has been recently 
proposed by theoretical design
as an optically active material in the 1.5 $\mu$m (0.8 eV) fiber optics
frequency window and with a monolithic integration with the Si (001)
surface.
Our results indicate that alloy formation in the group V planes would
likely occur at typical growth conditions. In addition, desired features
such as in-plane lattice constant and energy gap are virtually unchanged
and the optical oscillator strength for band-to-band transitions is
increased by a factor of 6 due to alloying.
\end{abstract}


\maketitle


\section{Introduction}
\label{Intro}

The monolithic integration between electronic and optically active materials
has been strongly pursued by the optoelectronics industry \cite{review}. 
Typical III-V and II-VI alloys, commonly used in the band gap engineering 
of optoelectronic devices, present a polarity mismatch \cite{harrison} for integration
with group-IV materials surfaces in the chemical bonding level
, i.e., heteroepitaxial depositions of group-V(III)
atoms followed by group-III(V) atoms on group-IV (001) surface produce an
excess (a lack) of electrons to make the chemical bonds. Moreover, none of 
these alloys do not satisfy 
constraints of having both lattice matching to silicon $and$ a direct band-gap in 
the 1.5 $\mu$m frequency range.

The growth sequence of group elements V-II-V-IV
proposed recently by Wang {\it et al.} \cite{wang1,wang2}
could solve all these problems.
The $(ZnSi)_{2}PAs_{3}$ compound was proposed after 
extensive theoretical exploration of several possible V-II-V-IV materials.
The new optoelectronic material could be grown using the heteroepitaxy
 sequence  As-Zn-As-Si-As-Zn-P-Si on a Si (001) surface, 
as shown in Fig. 1. 
This particular proportion between As and P optimizes the
lattice constant, the direct energy-gap and the polarity at the same time,
although a residual dipole moment is left due to chemical difference 
between P and As. Wang {\it et al.} \cite{wang1,wang2} 
proposed to eliminate this dipole inverting 
every other cell in the growth direction.

\begin{figure} 
\includegraphics[width=4.5in]{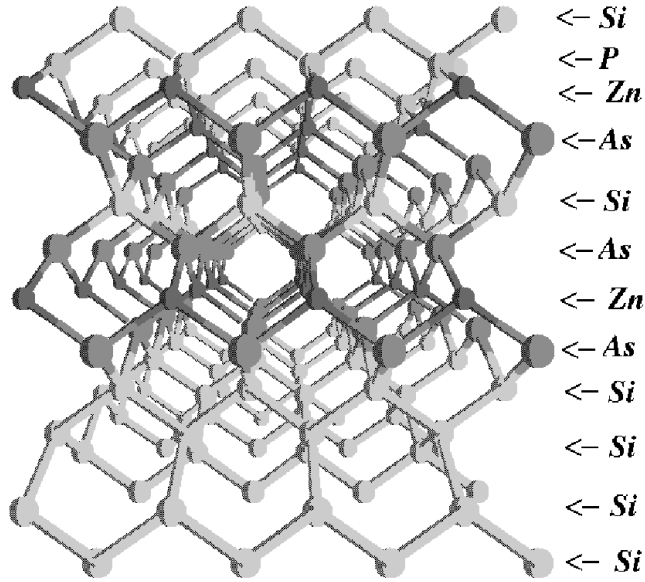}
\caption{Originally proposed $(ZnSi)_{2}PAs_{3}$ ordered material
on a Si (001) surface.}
\label{leitao_fig1}
\end{figure}

In this work, we explore the possibility of mixing As and P in 
the same growth plane in a 3:1 rate
instead of growing sequences of planes containing only As or only P.
Group-V mixing would automatically eliminate any residual dipole
moment. Besides, it could provide a simpler way to fine-tune the material's
properties (like the band gap and the lattice constant). Moreover, 
in real growth conditions, group-V 
mixing could be unavoidable due to the similar chemical characteristics
of P and As. Of course, it remains to be shown that the disordered
alloy retains the desired features of the ordered material, and this is the 
purpose of this work.

We perform {\it ab-initio} total-energy, band structure and optical
oscillator strength calculations based on Density Functional Theory (DFT)
and the pseudopotential method.
We compare the calculated 
properties of the originally designed
$(ZnSi)_{2}PAs_{3}$ and three forms that mix As and P in
the same epitaxial
growth plane. We also perform calculations on the "precursor" materials $ZnSiAs_2$,
$ZnSiP_2$ and $ZnSiAsP$ in order to 
understand the electronic structure features in the familly V-II-V-IV materials  
and their influence in the optical oscillator strength.

\section{Methodology}
\label{Model}

The present results have been obtained thanks to the ABINIT code \cite{gonze,ABINIT}, 
that is based on pseudopotentials and planewaves. It relies on an efficient Fast Fourier 
Transform algorithm \cite{FFT} for the conversion of wavefunctions between real and 
reciprocal space, on the adaptation to a fixed potential of the band-by-band conjugate 
gradient method \cite{payne} and on a potential-based conjugate-gradient algorithm 
for the determination of the self-consistent potential \cite{gonze2}. 

We use DFT within the local-density 
approximation (LDA) \cite{hohenberg,kohn}. We adopt
the exchange-correlation functional parametrized by Perdew and
Zunger \cite{perdew} from Ceperley and Alder's \cite{ceperley} data for correlation energy 
of the homogeneus electron gas. We use the scheme of Troullier 
and Martins \cite{tm} to generate soft norm conserving
pseudopotentials. The semi-local pseudopotentials are further transformed into fully 
separable Kleinman-Bylander pseudopotentials \cite{kleinman}, with the $d$ potential chosen as 
the local potential. For Zn atoms we use non-linear core corrections and an 
excitated and ionized configuration $s^{0.5}~p^{0.25}~d^{0.25}$. The pseudopotentials 
have been generated by the FHI98PP code \cite{FHI98PP}. 
The wave functions are expanded in a plane wave basis with maximum kinetic energy 
of 40 Ry. Brillouin-Zone summations are carried out in Monkhorst-Pack (4 x 4 x 4),
(4 x 4 x 2) and (2 x 2 x 4) $\vec k$-point sampling grid for precursors, 
$(ZnSi)_{2}PAs_{3}$ and alloys, respectively.

Calculations are performed for the precursors $ZnSiAs_2$, $ZnSiP_2$ and $ZnSiAsP$
(4 atoms per unit cell), the originally proposed 
$(ZnSi)_{2}PAs_{3}$ (eight atoms per unit cell)
and three distinct configurations of $(ZnSi)_{1/2}P_{1/4}As_{3/4}$ "alloys"
(16 atom per unit cell) that mix As and P in the same plane.  
The three alloy configurations (A-1, A-2 and A-3)
are shown in Fig. 2. The A-1 configuration has two planes with As:P 3:1, the A-2 and A-3
have one plane with four As and one plane with As:P 2:2 in two inequivalent
configurations. 

\begin{figure}
\includegraphics[width=4.5in]{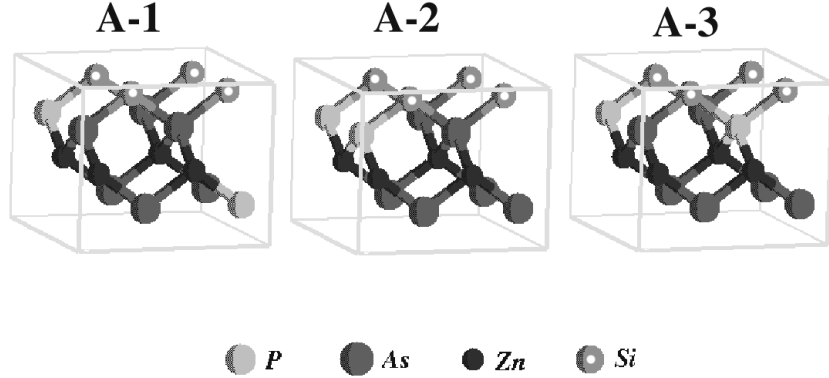}
\caption{ The three $(ZnSi)_{1/2}P_{1/4}As_{3/4}$ alloy
configurations using 16-atom unit cells.}
\label{leitao_fig2}
\end{figure}

As one can see, the A-2 and A-3 structures are not primitive, but we use the supercells
shown in Fig. 2 in order to facilitate comparisons among all three alloy structures.
Of course, only A-1 contains As:P 3:1 in each group-V
plane, as we propose.
But the other two are also important to study trends in the materials'
properties due to composition fluctuations.
The equilibrium lattice parameters for all structures
are found by minimizing the total energy.  For each set of lattice parameters 
the relative ion positions are relaxed until 
the forces are smaller then 10$^{-6}$ Hartree/Bohr.



\section{Results and Discussion}
\label{Result}

When fully relaxed, the alloy structure A-1 has a primitive monoclinic 
Bravais lattices (space group $P2_1$), but with very small deviations 
from a tetragonal unit cell: Differences in the in-plane lattice constants 
are smaller than 0.01 \AA { } ($a \simeq b =7.59$ \AA), and the angle between 
the respective lattice vectors is $90.005^o$. Structures A-2 and A-3 are 
primitive orthorhombic (space group $Pmm2$) and base-centered orthorhombic 
(space group Cmm2), respectively, with $a = 7.58$ \AA { } and $b = 7.60$ \AA { } 
in both cases. Therefore, they also present very small deviations from tetragonal 
($a=b$) symmetry. When considering epitaxial growth of disordered alloy structures
on top of a Si substrate, tetragonal symmetry will be very naturally imposed. 
Therefore, in the remainder of this article, we will adopt a tetragonal 
symmetry constraint.

\begin{table}
\caption{Lattice constants, in-plane lattice constant
mismatch with respect to Si and relative formation energies per group-V
atom ($\Delta E$) for the $(ZnSi)_{2}PAs_{3}$ compound and the A-1, A-2
and A-3 alloys. $\Delta E$ is calculated with respect to the precursor compounds
$ZnSiAs_2$ and $ZnSiP_2$.}
\begin{tabular}{ccccc}
Unit cell & $a$(\AA) & $c$ (\AA)  & mismatch (\%) & $\Delta E$ (eV) \\
\colrule
$(ZnSi)_{2}PAs_{3}$ & 3.81 & 10.91  & -0.2 & 0.001 \\
A-1 & 7.59 & 5.46 & -0.4 & -0.003 \\
A-2 & 7.59 & 5.46 & -0.4 & -0.002 \\
A-3 & 7.59 & 5.46 & -0.4 & -0.002 \\
\end{tabular}
\end{table}

Table I shows the calculated tetragonal lattice constants ($a$ and $c$), 
the in-plane lattice mismatch with respect to Si and the formation energies for 
the $(ZnSi)_{2}PAs_{3}$ compound and the A-1, A-2 and A-3 alloys.
The mismatch is
$(a - a_{Si})/a_{Si}$ for unmixed $(ZnSi)_2P_{1/4}As_{3/4}$ and
$(a - 2a_{Si})/2a_{Si}$ for the alloys, where $a$ is the in-plane lattice
constant for the new materials and $a_{Si}$ is the Si lattice constant
multiplied by $\sqrt{2}/2$. 
Our results show that the in-plane mismatch for all alloy structures
are very small, suggesting that disorder in the group-V planes leads
to structures that can be monolithically integrated to Si as easily as 
the originally proposed ordered material.

\begin{figure}
\includegraphics[width=4.5in]{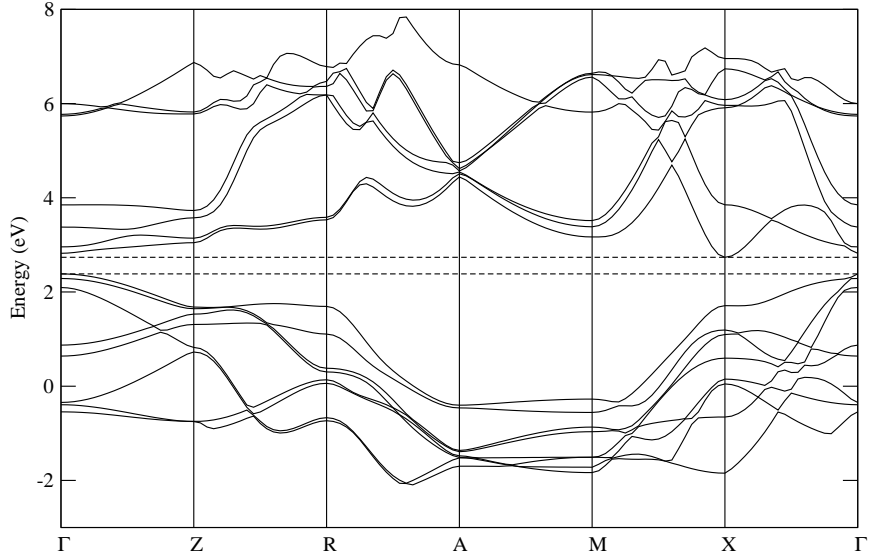}
\caption{LDA band structure of the $(ZnSi)_{2}PAs_{3}$ compound.
The dashed lines shows the gap region. }
\label{leitao_fig3}
\end{figure}

The relative formation energies per group-V atom for the alloys and the 
$(ZnSi)_{2}PAs_{3}$ compound are calculated with respect to the precursor
compounds $ZnSiAs_2$ and $ZnSiP_2$, i.e., they define the enthalpy of mixing
different group-V elements. As one can see, all values are extremely small,
slightly negative for the alloys (from -2 to -3 meV) and slightly positive for
$(ZnSi)_2PAs_3$. These small values allow us to safely state 
that structures with the same proportion of As and P are 
essentially degenerate in energy, regardless of the particular atomic arrangement.
Moreover, this suggests that, at typical growth temperatures, disordered structures
will have lower free energy due to their higher configurational entropy.  

\begin{figure}
\includegraphics[width=4.5in]{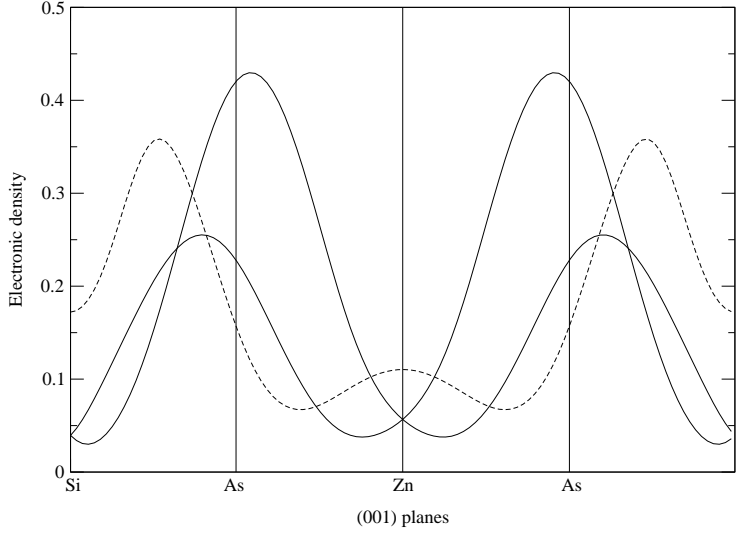}
\caption{Electronic density in the (001) planes
of $ZnSiAs_2$  Full lines correspond to the two degenerate states at the
top of the valence band and the dashed line corresponds to the
state at the bottom of the conduction band.}
\label{leitao_fig4}
\end{figure}

We now focus on the group-V mixing effects in the band structure of these
materials. Fig. 3 shows the band structure of the 
 $(ZnSi)_{2}PAs_{3}$ compound. Within LDA this compound 
has a marginally indirect gap,
with valence band maximum at $\Gamma$ and conduction band minimum at $X$.
Quasi-particle $GW$ corrections make it a direct-gap material with a $\sim$ 0.8 eV
energy gap at $\Gamma$ \cite{wang1,wang2}. 
Table II shows $\Gamma \rightarrow \Gamma$ and $\Gamma \rightarrow X$
energy gaps for the $(ZnSi)_{2}PAs_{3}$ 
compound and the three alloy
configurations. Clearly, the energy gaps at $\Gamma$ and $X$ are not affected too
much by group-V mixing \cite{comment}. Moreover, the LDA "indirectness" 
(measured by the ($\Gamma \rightarrow \Gamma$) - ($\Gamma \rightarrow X$) 
gap difference)
is also essentially unaffected. From these considerations, it is likely that $GW$
corrections will also be similar in all these structures, and we may expect that
alloys with As:P 3:1 proportion will also have a direct band-gap around 0.8 eV.
More important, the tayloring of structures with group-V 
mixing provides an easier way to "fine-tune" the
the gap energy so that matches exactly the 1.5 $\mu$m fiber optics
window.

\begin{table}
\caption{LDA energy gap (eV) and relative OOS for the $(ZnSi)_{2}PAs_{3}$
compound and the A-1, A-2 and A-3 alloys.}
\begin{tabular}{ccccc}
Unit cell & $\Gamma \rightarrow \Gamma$ & $\Gamma \rightarrow X$ &
$(\Gamma \rightarrow \Gamma) - (\Gamma \rightarrow X)$ &
 $f_{vc}$ $(\Gamma \rightarrow \Gamma)$  \\
\colrule
$(ZnSi)_{2}PAs_{3}$ & 0.438 & 0.353  & 0.085 & 1.00 \\
A-1 & 0.552 & 0.463 & 0.089 & 6.04 \\
A-2 & 0.498 & 0.393 & 0.105 & 1.22 \\
A-3 & 0.523 & 0.391 & 0.133 & 1.16 \\
\end{tabular}
\end{table}

\begin{figure}
\includegraphics[width=4.5in]{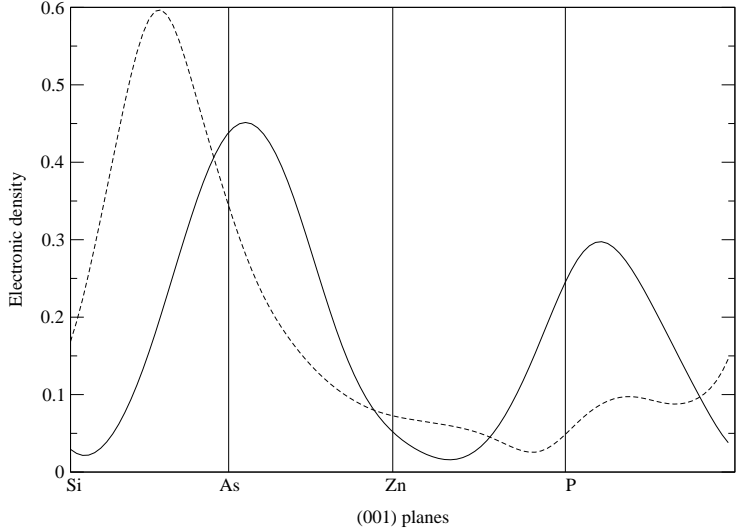}
\caption{Electronic density in the (001) planes
of $ZnSiAsP$  Full lines correspond to the top of the valence band and the
dashed line corresponds to the bottom of the conduction band.}
\label{leitao_fig5}
\end{figure}

\begin{figure}
\includegraphics[width=4.5in]{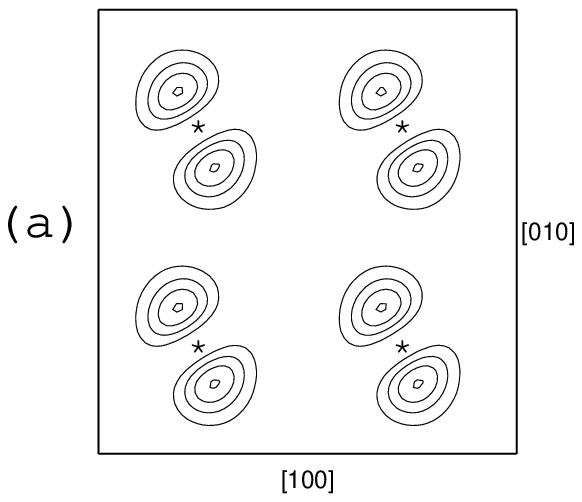}
\includegraphics[width=4.5in]{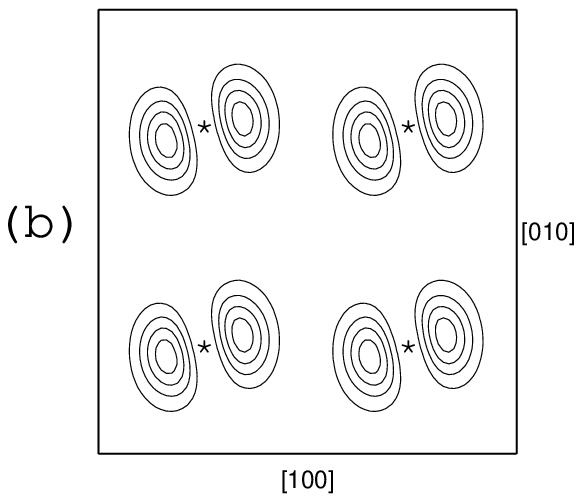}
\caption{Valence band electronic densities in the As planes for the
precursor $ZnSiAs_2$ for the two degenerate states ((a) and (b))
at the top of the valence band. The stars show the As nuclei positions
The frame is 8\AA $\times$ 8\AA ~and the contours run from 2 to 8
electrons/\AA$^3$.}
\label{leitao_fig6}
\end{figure}

\begin{figure}
\includegraphics[width=4.5in]{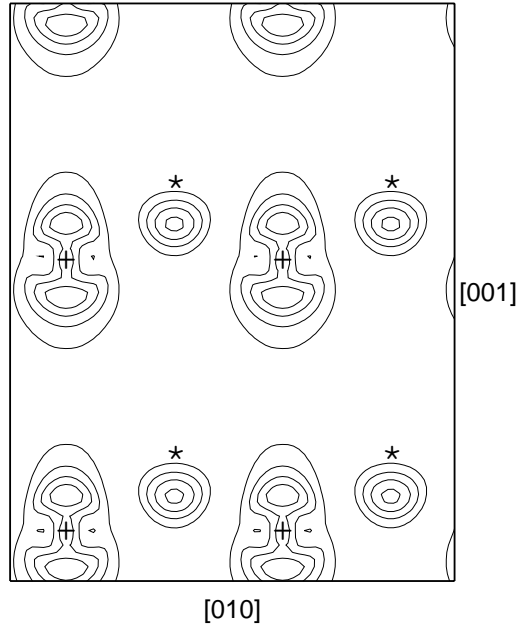}
\caption{Conduction band electronic densities in the (011) plane
for the precursor $ZnSiAs_2$. The stars and the crosses show the As
and Si nuclei positions, respectively. The frame is 8\AA $\times$ 12\AA
~and the contours run from 2 to 8 electrons/\AA$^3$.}
\label{leitao_fig7}
\end{figure}

\begin{figure}
\includegraphics[width=4.5in]{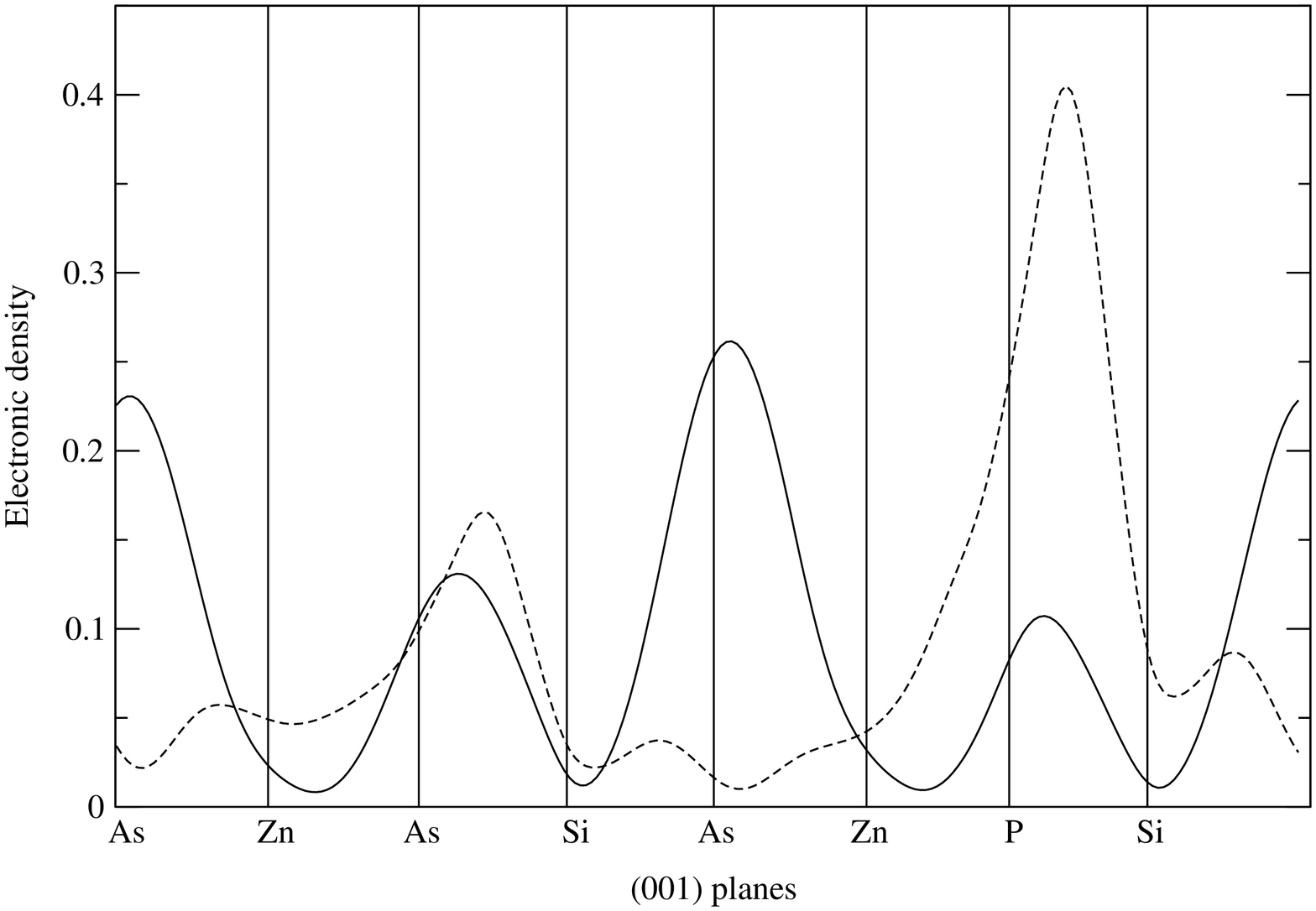}
\caption{ Electronic density in the (001) planes
for the $(ZnSi)_{2}PAs_{3}$ compound.  The
full line is the valence band and the dashed line is the conduction
band.}
\label{leitao_fig8}
\end{figure}

\begin{figure}
\includegraphics[width=4.5in]{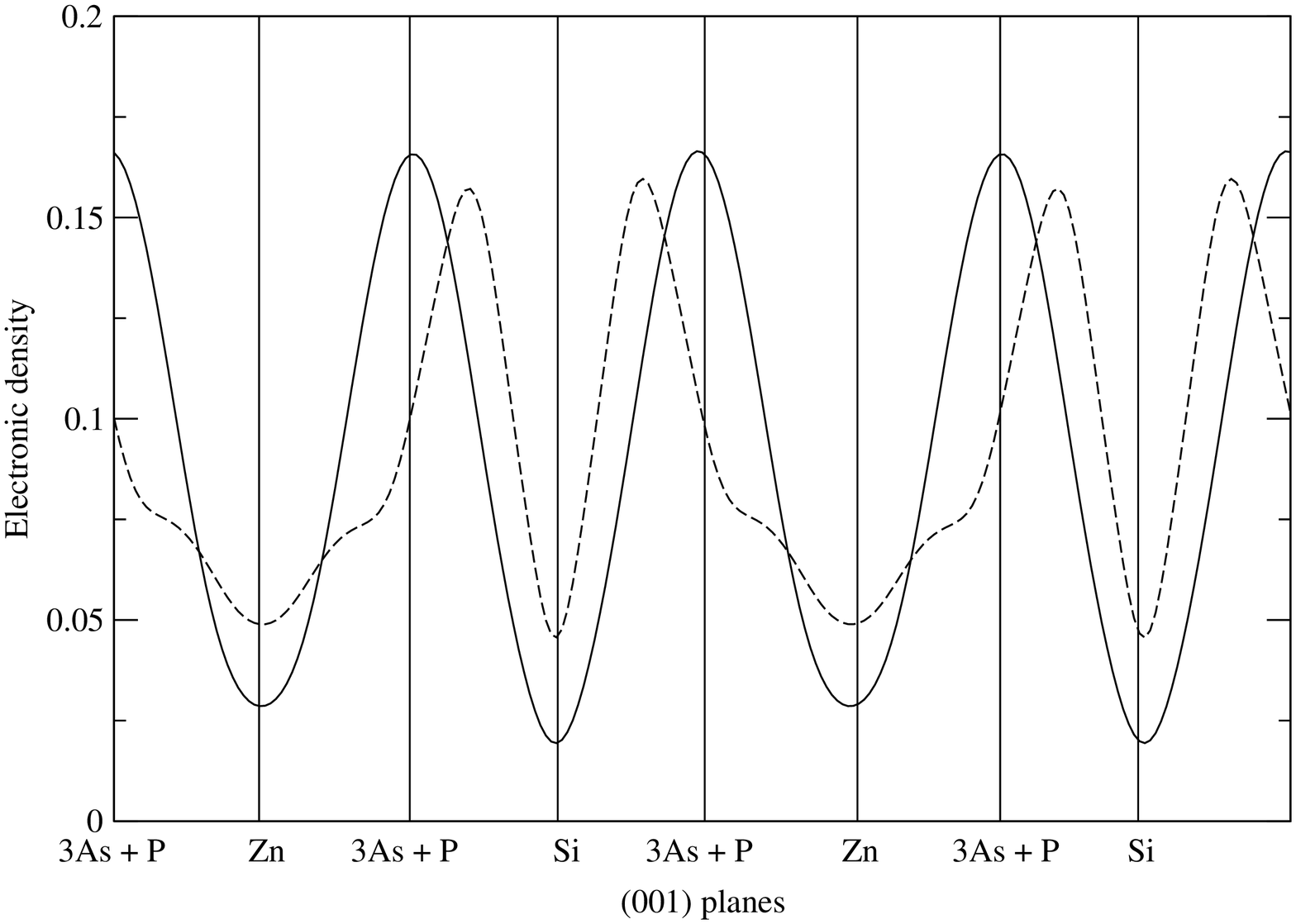}
\caption{Electronic density in the (001) planes for the A-1 alloy configuration.
The full line is the valence band and the dashed line is the conduction
band.}
\label{leitao_fig9}
\end{figure}

\begin{figure}
\includegraphics[width=4.5in]{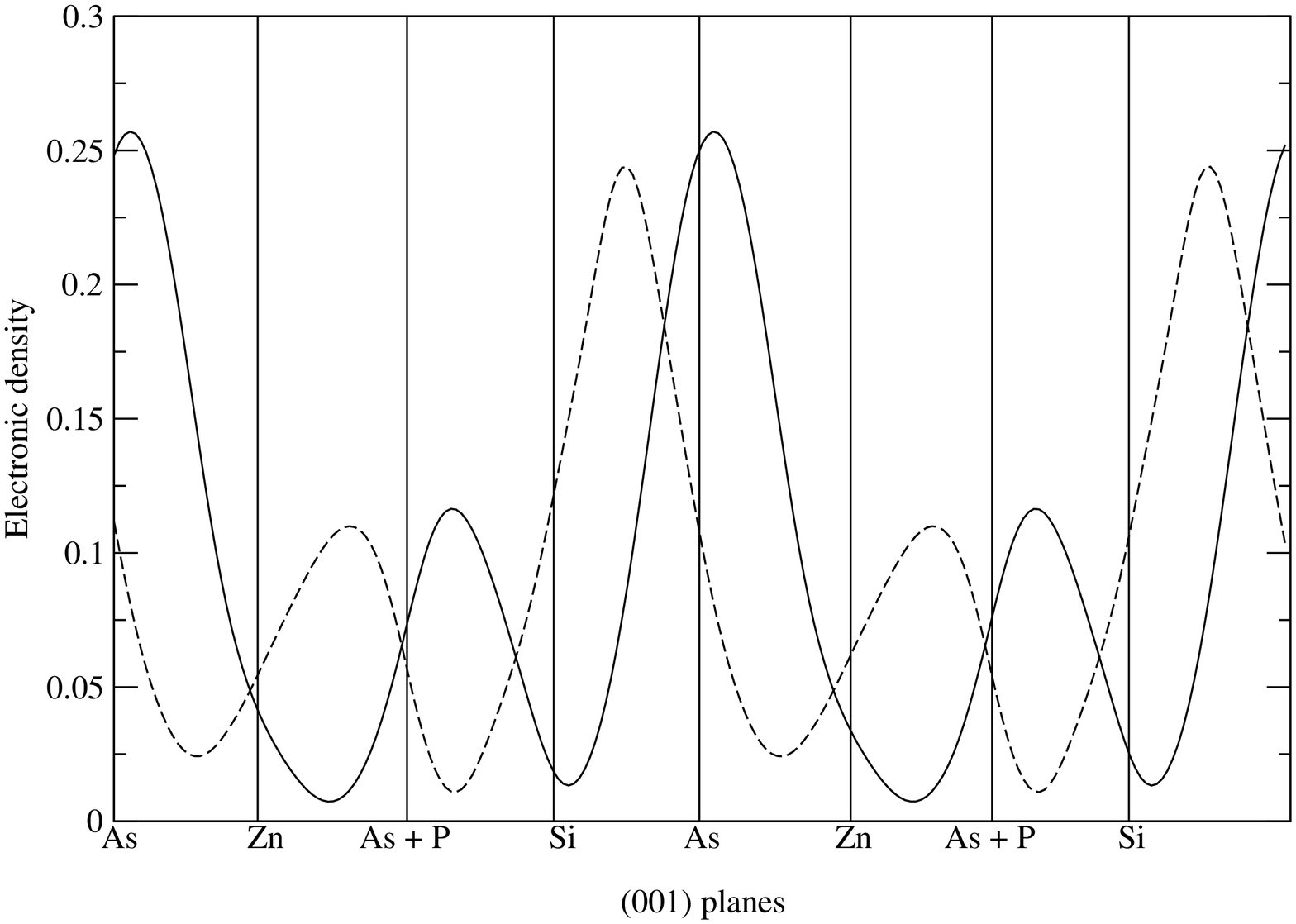}
\caption{Total electronic density in the (001) planes for the A-3 alloy configuration.
The full line is the valence band and the dashed line is the conduction
band.}
\label{leitao_fig10}
\end{figure}

A direct gap is not enough to guarantee efficient emission properties.
In addition, one has to ensure a high optical oscillator strenght (OOS)
between the initial and final states. Table II also shows the calculated 
OOS for conduction to valence transitions
at $\Gamma$, given by: 
\begin{equation}
f_{vc} = {\frac{2}{3m}}{\frac {|{\langle v | \vec P | c  \rangle}|^2}{E_c - E_v}} ,
\label{explicit2}
\end{equation}
where $\vec P$ is the
momentum, $E_c - E_v$ is the gap energy $\Gamma$ and $m$ is the mass of
the electron. The values are normalized by $f_{vc}$ for the $(ZnSi)_{2}PAs_{3}$ 
compound. The results show
that the $(ZnSi)_{1/2}P_{1/4}As_{3/4}$ alloy with As:P 3:1 in each
plane containing V-atoms (A-1) is the best emitter. Comparing with the 
$(ZnSi)_{2}PAs_{3}$, the OOS is multiplied by six.
The configurations A-2 and A-3 have higher OOS's than $(ZnSi)_{2}PAs_{3}$
too, but only by 16\% and 22\%, respectively.
This result is striking, because in general one naively expects that disorder
tends to decrease OOS's, but we shall see that it can be 
explained from a simple analysis of 
the electronic density profiles of valence and 
conduction states for the structures.

First of all, it is instructive to analyze the density profiles
of band-edge states
for the precursors $ZnSiAs_2$ and $ZnSiAsP$ in the Figs. 4 and 5, respectively.
These figures show the projected electronic density on the growth axis
(i. e., averaged over the $xy$ planes).
The top of the valence band in the precursor $ZnSiAs_2$ is doubly
degenerated, with most of the charge in the As planes as shown in Fig. 4.
In fact, the electrons in these bands are mainly in linear combinations of
$p_x$ and $p_y$ atomic orbitals as shown in Fig. 6. 
Fig. 4 also shows that
the density from the conduction band of $ZnSiAs_2$ is peaked
between Si and As atomic planes. Indeed, it is
formed by antibonding states between As and Si (Fig. 7).

When one of the As planes is replaced by a P plane  to form $ZnSiAsP$, 
the degeneracy in the top of the valence band is broken and the symmetry of conduction
electronic density too, as show in Fig. 5. The valence band charge is now
mostly on the As planes and the conduction band one is between Si and As 
planes.


The density profiles for the $(ZnSi)_{2}PAs_{3}$ compound and related alloys
seem to obey the precursors' general scheme. 
None of these structures has energy degeneracy in the top of the valence band.
The valence band electrons are mainly near the group-V planes and the conduction
band electrons are antibonding states between Si-As or Si-P, for all
configurations. 
For $(ZnSi)_{2}PAs_{3}$ (Fig. 8), the charge density maxima
at the valence band alternate their intensities like in the precursor $ZnSiAsP$, 
but the conduction band electrons are mostly located between 
the Si and P planes. There is a significant 
difference between A-1 and the ordered $(ZnSi)_{2}PAs_{3}$, since all group-V planes
are identical. Because of that, the valence band 
electronic densities at these planes are all the same, as shown in Fig. 9.
The density of the antibonding states at conduction band is also evenly 
distributed between all 
equivalent group-V planes and Si planes. We can clearly note that the overlap
between top valence states and the conduction states in the A-1
alloy is much larger than originally proposed  $(ZnSi)_{2}PAs_{3}$ compound.
That is the reason why the OOS for the A-1 alloy is 
much higher than that of the $(ZnSi)_{2}PAs_{3}$ compound.

For the alloy configuration A-3, the densities of 
valence and conduction charge (Fig. 10)
clearly show an intermediate overlap if compared to 
A-1 and $(ZnSi)_{2}PAs_{3}$. The situation
is very similar for the configuration A-2. From all these 
results, we can extract the following
trend for valence-conduction overlap in real space: A-1 $>>$
 A-3 $\approx$ A-2 $>$ $(ZnSi)_{2}PAs_{3}$, which, therefore 
explains the trend in OOS's shown in Table II . As a general statement, 
we can say that a more homogeneous distribution 
between As and P atoms in the group-V planes is the key to 
maximize the emission properties of
this class of materials. \cite{footnote}

\section{Conclusions}
\label{concl}

In conclusion, we have calculated the total energy, geometry, optical oscillator strength,
band structure and electronic densities profiles of $(ZnSi)_{1/2}P_{1/4}As_{3/4}$
as an ordered material and in three alloy configurations in order to 
study the group-V mixing effect on
the properties of this new optoelectronic material.
Our results indicate that the relevant structural (in-plane lattice matching to Si)
and optical (band gap at $\sim$ 0.8 eV) properties are kept in the presence 
of group-V atomic mixing.
Very small energy changes between different configurations suggests 
that, at growth temperatures, 
random occupation of group-V planes would minimize the free energy. 
The mixing would allow fine-tuning of the
band gap energy. The calculated optical oscillator strengths suggest
that the structure with As and P at 3:1 rate in all group-V planes would be
six times more optically active than the structure originally proposed, with As and P in 
different (001) planes.

\acknowledgments
This work was supported by the following Brazilian funding agencies:  
Conselho Nacional de Desenvolvimento Cient\'\i fico e
Tecnol\'ogico (CNPq), Funda\c c\~ao Carlos Chagas Filho de Amparo \`a 
Pesquisa do Estado do Rio de Janeiro (FAPERJ),
Funda\c c\~ao Universit\'aria Jos\'e Bonif\'acio (FUJB-UFRJ),
Instituto do Mil\^enio de Nanoci\^encias and 
Programa de N\'ucleos de Excel\^encia  (PRONEX-MCT).


\begin{thebibliography}{10}
\bibitem{comment}
It is important to notice that 0.1 eV energy differences are within
the error bars of either LDA and GW band structures.
\end{thebibliography}
\end{document}